\documentstyle[11pt,aas2pp4]{article}

\lefthead{Bell et al.}

\righthead{HC$_{11}$N in TMC-1}

\begin{document}

\def\vs{\vskip 0.3 true cm}

~
\vs
\vs
\vs

\title{Detection of HC$_{11}$N in the Cold Dust Cloud TMC-1}
                                    
\author{M.B. Bell\altaffilmark{1}, and P.A.
Feldman\altaffilmark{4} \nl and \nl M.J. Travers\altaffilmark{2,3},
M.C. McCarthy\altaffilmark{2,3}, C.A. Gottlieb\altaffilmark{3},
and  P. Thaddeus\altaffilmark{2,3}}

\altaffiltext{1}{Herzberg Institute of Astrophysics,  National
Research Council of Canada, 100 Sussex Drive, Ottawa, Canada K1A
0R6}
\altaffiltext{2}{Harvard-Smithsonian Center for Astrophysics,  60
Garden Street, Cambridge, MA 02138}
\altaffiltext{3}{Division of Engineering and Applied Sciences, 
Harvard University, 29 Oxford  Street, Cambridge, MA 02138}
\altaffiltext{4}{DAO, Herzberg Institute of Astrophysics, NRC of
Canada,  5071 W.  Saanich Road, Victoria, BC, Canada V8X 4M6.}

\begin{abstract} Two consecutive rotational transitions of the long
cyanopolyyne HC$_{11}$N, $J=39 \rightarrow 38$ and $38 \rightarrow
37$, have been detected in the cold dust cloud TMC-1 at the 
frequencies expected from recent laboratory measurements by
Travers et al. (1996), and at about the expected intensities. The
astronomical lines have  a mean radial velocity of 5.8(1) km
s$^{-1}$, in good agreement with the shorter cyanopolyynes
HC$_{7}$N and HC$_{9}$N observed in this very sharp-line source
[$5.82(5)$ and $5.83(5)$ km s$^{-1}$, respectively].  The column
density of HC$_{11}$N is calculated to be $2.8\times10^{11}$
cm$^{-2}$.  The abundance of the cyanopolyynes decreases smoothly
with length to HC$_{11}$N, the decrement from one to the next
being about 6 for the longer carbon chains.
\end{abstract}


\twocolumn    

\section{Introduction} Several weak radio lines in the
circumstellar shell of  IRC+10216  (Bell et al. 1982) and the dust
cloud TMC-1 (Bell \& Matthews 1985) were originally  attributed to
the long carbon-chain molecule HC$_{11}$N (cyano-deca-penta-yne),
but a sensitive spectral survey of IRC+10216 between
$22.3$ and $24.1$ GHz  (Bell unpublished) showed no evidence for
this molecule at other frequencies predicted from the earlier
assignment.  This anomaly has now been explained with the
laboratory detection of HC$_{11}$N by means of a Fourier transform
molecular-beam spectrometer  which showed that the  earlier
assignment was incorrect (Travers et al. 1996).  Although the deep
IRC+10216 spectral survey also showed no evidence for spectral
lines at the HC$_{11}$N laboratory frequencies (Bell, Feldman, \&
Avery 1992; Bell et al. 1992), the laboratory data have permitted
a very deep astronomical search for this molecule in the
narrow-line source TMC-1, and this has now been done with the new,
very sensitive, dual-polarization GBT receiver in operation on the
NRAO\footnote{The National Radio Astronomy Observatory (NRAO) is
operated by Associated Universities Inc., under agreement with the
National Science Foundation.} 140-foot (43 m) telescope. We report
here the detection in TMC-1 of two transitions of HC$_{11}$N near
13 GHz, close to the expected line intensity peak for a source at
10 K.  

\section{Observations and Results} \label{hairymath}

The observations were done in two sessions with the 140-foot
telescope, both during excellent weather.  In 1996 December the
$J=39 \rightarrow 38$ line  at 13186 MHz was observed for a total
of 35 hours and in 1997 January the
$38 \rightarrow 37$ line at 12848 MHz was observed for a total of
36 hours. The observations during the second session were
undertaken remotely from Ottawa by means of the 140-foot DISPLAY
software.

Telescope pointing difficulties encountered in both sessions
sometimes  required pointing checks every 30 minutes. For this
reason, during most of each session a nearby transition of
HC$_{9}$N was observed simultaneously in the second correlator
bank, which is available for independent observations even when
the dual-polarization receiver is used (provided that the 
autocorrelator is set in the 4-bank mode). In both sessions, on
starting  observations each day the $J=5 \rightarrow 4$ transition
of HC$_{5}$N at 13313 MHz and the $12 \rightarrow 11$ transition
of HC$_{7}$N at 13536 MHz were also measured. To calculate
accurate line intensities we also observed HC$_{5}$N and HC$_{7}$N
near the meridian, so telescope gain changes could be estimated
and corrections applied during data reduction. 

For the HC$_{11}$N observations, three different local-oscillator
(LO) frequencies were used over the two sessions so that features
that moved with the LO frequency could be recognized.  With the
narrow line widths in TMC-1 ($0.5$ km s$^{-1}$, or 20 kHz at 13
GHz), spectral features of terrestrial origin are smeared over
several line widths during the 4-5 day observing sessions.  Sharp
features that are present in each night's average, or in both the
first and last halves of the data, must then be either
source-related or an artifact introduced by the correlator.  Since
the HC$_{11}$N spectra were in banks 2 and 4 in December and banks
1 and 3 in January, U~lines that occur in different channels in
the two sessions might still be correlator related.  Because of
the change of banks between the two sessions, however, the
appearance of the HC$_{11}$N line at the expected position in both
sessions is highly unlikely to be of instrumental origin. Although
spectral baseline ripple is not usually a problem when frequency
switching with small   offsets ($\pm29$ kHz) on narrow-line
sources, the normal $\pm\lambda/8$ focus modulation that removes
ripple produced by  residual standing waves between the primary
and secondary reflectors was also employed as a precaution.

In December the observed frequencies were near the center of the
receiver passband and system temperature was about 28 K; in
January they were closer to the edge of the band and the system
temperature was about 33 K.  The telescope beamwidth (FWHM) was
$2.'4$ and the beam efficiency  was $0.54$ (C. Brockway, personal
communication 1992).  The observing technique used was
small-offset frequency switching; the frequency-switched images
were later removed as described  previously (Bell, Avery, \&
Watson 1993; Bell 1996; Bell 1997).  The autocorrelator was
configured with 4 banks  (2 for each polarization), each with 256
channels.  The frequency offset used for HC$_{9}$N and HC$_{11}$N
was $\pm6$ channels.  An offset of $\pm16$ channels was used for
HC$_{5}$N and HC$_{7}$N because of the hyperfine splitting of the
HC$_{5}$N transition.  Taking both polarizations into  account,
the effective on-source integration time for HC$_{11}$N was about 
70~h in each session. 

Spectra containing the two observed transitions of  HC$_{11}$N,
and the 
$J=23 \rightarrow 22$ transition of HC$_{9}$N observed
simultaneously with the
$39 \rightarrow 38$ transition of HC$_{11}$N, are presented in
Figure~1 plotted on a common velocity scale.  All three spectra
have been processed in an identical fashion, which included
overlapping, removing the reference images, and Hanning
smoothing.  The line parameters are presented in Table 1. The rest
frequency obtained from laboratory work is included in column 2,
the measured LSR velocity is in column 3, the brightness
temperature is in column 4, and the measured line width (FWHM) is
in column 5.  As can be seen from column 3, the velocities
measured for HC$_{7}$N and HC$_{9}$N are close to the value of
$5.85$ km s$^{-1}$ expected for this source (Kroto et al. 1978; 
T\"{o}lle et al.\ 1981).  Although the velocities of the
HC$_{11}$N lines have larger uncertainties because of the poorer
signal to noise, the mean value obtained from the two transitions
($5.84$ km s$^{-1}$) is in excellent agreement with the values
found for the shorter cyanopolynes.  The lines identified as
HC$_{11}$N also have widths normal for TMC-1.

It should be noted that in some cases the observing frequency used
differed slightly  from the rest frequency listed in column 1 of
Table 1.  For  HC$_{5}$N the observing frequency used was
$13313.338$ MHz.  More importantly, for the HC$_{9}$N $J=22
\rightarrow 21$ and $23
\rightarrow 22$ lines the observing frequencies were $12782.766$
MHz and $13363.790$ MHz, respectively,  and the differences
between these and the rest frequencies in Table 1 were taken into
account in determining velocities.  For the $J=38 \rightarrow 37$
line of HC$_{11}$N, the value obtained from the laboratory
constants, $12848.731$ MHz, was used both as an observing
frequency and in determining the velocity listed in Table 1.

Three U~lines are apparently present in the deep HC$_{11}$N
spectra in Figure~1.  For the reason mentioned earlier we are
unable to rule out completely the possibility that these are
spurious features introduced by the correlator.  If this is the
case, however, they are seen only at very low levels.  From the
two HC$_{11}$N spectra, the line density at this level ($\sim2$mK)
is estimated to be $\sim2$ lines per MHz.  It is easily shown that
with this line density the chance of detecting a  U~line within
one channel (4.88 kHz) of the expected HC$_{11}$N position is
about 1\%, and the probability of this happening for two
transitions is thus very small: 
$\sim10^{-4}$.  The probability of finding lines at the right
frequencies with the right intensities and widths is clearly even
less, so the identification of HC$_{11}$N in space now seems quite
secure.

Spectra obtained for HC$_{5}$N and HC$_{7}$N (after 4 and 20
minutes of  integration, respectively) are shown in Figure~2 after
overlapping.  In these  cases overlapping has been done but their
frequency-switched images have not been removed.  To determine the
integrated line intensity of HC$_{5}$N, the area under the three
quadrupole hyperfine components was summed.  To obtain the final
single-component peak equivalent value, the total area was then
divided by $0.4$ km s$^{-1}$ which was assumed to be the velocity
width (FWHM) of the source (Broten et al. 1978). 

Column density estimates were made for all four observed
cyanopolyyne  molecules on  the assumption that (i) all lines are
optically thin, (ii) all lines originate from a source of size
$6.'0$ by $1.'3$ (Churchwell et al. 1978; Olano et al. 1988), and
(iii) the  excitation temperatures are identical (i.e., there is a
single rotational temperature).  Calculations were carried out for
rotational temperatures (T$_{rot}$) of 6, 8, 10, and 12 K,
following the analysis of Avery et al. (1992).  Although it might
be thought that molecules with the very large dipole moments ($>4$
D) of the present cyanopolyynes would cool quickly and have
rotational excitation temperatures lower than the kinetic
temperature of the gas, this is not expected to be the case for
the low-lying transitions observed here.  Since the Einstein A
coefficients are proportional to $\nu^{3}\mu^{2}$, the radiative
cooling rates remain small even for transitions as   high as
$J=40$, and the excitation temperatures are therefore expected to
be  close to the kinetic temperature.  If so, the excitation
temperature of $\sim11.5$ K recently determined for HC$_{9}$N in
TMC-1 by Bell et al. (1997) is probably a  good measure of the
kinetic temperature of the gas in this source (c.f. Dickman 1975;
Avery 1980).

The dipole moments used in the excitation analyses are listed in
Table 2 along with  the column densities calculated for T$_{rot}$
= 10 K.  The dipole moments  quoted by Snyder et al. (1986) for
HC$_{7}$N, HC$_{9}$N and HC$_{11}$N, which were based on crude
ab-initio self-consistent field (SCF) calculations and modified by
making empirical adjustments and extrapolation, are probably too
low by as much as 0.5 Debye.  Recently, large-scale coupled
cluster [CCSD(T)] calculations have been carried out for
HC$_{7}$N, HC$_{9}$N, and HC$_{11}$N by P. Botschwina (personal
communication 1997).  He finds
$\mu_0$(HC$_{7}$N) = 4.82(5) D, $\mu_0$(HC$_{9}$N)=5.20(5) D, and
$\mu_0$(HC$_{11}$N)=5.47(5) D.  These values are estimated to be
accurate to within a few percent, even though HC$_{7}$N to
HC$_{11}$N fall in the strongly non-linear part of the
dipole-moment curve with respect to carbon-chain length.  The
column density found here for  HC$_{9}$N, $1.9\times10^{12}$
cm$^{-2}$, is slightly lower than the value $3.2\times10^{12}$
cm$^{-2}$ reported by Broten et al. (1978), although the dipole
moment of this molecule turns out to be slightly lower ($\sim7\%$)
than their estimate of 5.6~D.  This is largely due to the higher
antenna temperature that Broten et al. reported for the $25
\rightarrow 24$ transition at 14535~MHz, which is almost 50\%
higher than we would now predict. 
     
The calculated column densities are plotted for all four molecules
in Figure~3 (upper), for excitation temperatures of 6, 8, 10, and
12 K.  As the Figure shows, the abundance of HC$_{11}$N does not
vary significantly for excitation temperatures between 8 and 12
K.  Figure~3 (upper) also shows how the abundance changes with the
length of the cyanopolyyne chain, and how the decrements in
abundance steepen as the excitation temperature increases.  This
can be seen more clearly in Figure~3 (lower) where, for T$_{rot}$
= 10 K, the decrements are slightly larger than the value of
$\sim4$ found previously by Broten et al. (1978) ---  they
approach the earlier value, however, when lower excitation
temperatures are assumed. 

The present detection of HC$_{11}$N, a molecule with molecular
weight (147 amu) nearly twice that of glycine, demonstrates that
detectable quantities of fairly large organic molecules can form
in astronomical sources under conditions markedly different from
those on Earth.  However, if the column densities of these carbon
chains continue to drop at the rate observed for the longest
cyanopolyynes, the most intense lines of the next longer
cyanopolyyne, HC$_{13}$N, are unlikely to be stronger than $\sim1$
mK.  If so, detection of this molecule in TMC-1 is likely to
require the smaller, more efficient beam of the 100~m Green Bank
Telescope now under construction by NRAO.  

We wish to thank George Liptak and the 140-foot telescope
operators for their assistance, especially during the remote
observing session.  We thank Ron  Maddalena for providing the very
useful DISPLAY software package required  to carry out the remote
observations.  One of us (MBB) also thanks Pierre  Brault for his
assistance in writing the LINECLEAN software and Jim Watson for
helpful information and discussions on molecular physics.


\newpage

\begin{planotable}{lllrl}
\tablecaption{Line Parameters in TMC-1\tablenotemark{a}}
\tablewidth{0pt}
\tablehead{
\colhead{Molecule} & \colhead{Rest Frequency} & \colhead{$v_{\rm
LSR}$} & 
\colhead{$T_{\rm B}$\tablenotemark{b}} & \colhead{FWHM}  \nl
\colhead{(Transition)} & \colhead{(MHz)} & \colhead{(km s$^{-1}$)}
& \colhead{(mK)} & \colhead{(km s$^{-1}$)}  \nl}
\startdata HC$_5$N ($J=5 \rightarrow 4$)    &
~~~13313.3\tablenotemark{c}    &~~~------        & 1770(43)   &
~~0.4\tablenotemark{d}   \nl  HC$_7$N ($12 \rightarrow 11$)    &
~~~13535.991\tablenotemark{e}  & ~~5.82(5)  & 475(13)   
&~~0.53(2)   \nl  HC$_9$N ($22 \rightarrow 21$)    &
~~~12782.769\tablenotemark{f}  & ~~5.84(5)\tablenotemark{g}  &
81.3(15)  & ~~0.40(1)   \nl   HC$_9$N ($23 \rightarrow 22$)    &
~~~13363.800\tablenotemark{f}  & ~~5.84(5)\tablenotemark{g}  &
82.2(22)  &~~0.38(1)   \nl  HC$_{11}$N ($38 \rightarrow 37$) &
~~~12848.728\tablenotemark{h}  & ~~5.73(10)  & 9.4(17)   &
~~0.36(10)   \nl  HC$_{11}$N ($39 \rightarrow 38$) &
~~~13186.853\tablenotemark{h}  & ~~5.96(10) & 5.4(7)   &
~~0.56(10)   \nl U(13186.46)                      &             &
~~5.85           & 4.8(7)    & ~~0.26(10)  \nl 
U(13186.98)                      &             & ~~5.85          
& 6.3(11)   & ~~0.26(7)   \nl  U(12848.48)                     
&             & ~~5.85           & 7.4(20)   & ~~0.26(6)   \nl
\enddata
\tablenotetext{a}{$\alpha _{1950} = 04^{\rm h} ~ 38^{\rm m} ~
39.3^{\rm s}$, $\delta _{1950} = 25^\circ ~ 35^{\rm '} ~ 36^{\rm
''}$}
\tablenotetext{b}{$T_{\rm B} = (T_{\rm A}^*) / \eta_B$
(uncorrected for beam dilution)}
\tablenotetext{c}{From Alexander et al. 1976}
\tablenotetext{d}{Assumed value}
\tablenotetext{e}{From Kroto et al. 1978}
\tablenotetext{f}{From Travers 1996 (unpublished)}
\tablenotetext{g}{After adjustment for the observing frequency
used (see text)}
\tablenotetext{h}{From Travers et al. 1996}
\end{planotable}


\newpage
\begin{planotable}{llcc}
\tablecaption{Cyanopolyyne Column Densities in TMC-1 For $T_{\rm
rot} =10$ K}
\tablewidth{0pt}
\tablehead{
\colhead{Molecule} & \colhead{Dipole Moment ($\mu_{0}$)} &
\colhead{$N_L$ (10$^{11}$  cm$^{-2}$)\tablenotemark{a}} &
\colhead{Abundance} \nl
\colhead{} & \colhead{(Debye)} & \colhead{(cm$^{-2}$)} &
\colhead{Decrement}}
\startdata HC$_5$N        & ~~~~~~4.33\tablenotemark{b}    & 
330      &            \nl HC$_7$N        &
~~~~~~4.82\tablenotemark{c}    &  110      &     3.0    \nl
HC$_9$N        & ~~~~~~5.20\tablenotemark{c}    &   19      &    
5.8    \nl HC$_{11}$N     & ~~~~~~5.47\tablenotemark{c}    & 
2.8      &     6.8    \nl
\enddata
\tablenotetext{a}{Adopting a beam-dilution factor of
$(1.'3/2.'4)$= 0.54  (because the source is extended in one
dimension; see Broten et al. 1978)}
\tablenotetext{b}{From Alexander et al. 1976}
\tablenotetext{c}{From P. Botschwina 1997 (personal communication)}
\end{planotable}


\onecolumn

\clearpage

\figcaption[sgi9259.eps]{Spectra in TMC-1 of two lines of
HC$_{11}$N above one of  HC$_{9}$N.  Obtained by small-offset
frequency switching, the spectra have been overlapped and the
frequency-switched images removed.
\label{fig1}}

\figcaption[sgi9259.eps]{Spectra (upper) of the $J=5 \rightarrow
4$ transition of HC$_{5}$N, showing partially resolved quadrupole
hyperfine structure, and  (lower) of the
$12 \rightarrow 11$ transition of HC$_{7}$N.  The
frequency-switched images have not been removed. 
\label{fig2}}

\figcaption[sgi9259.eps]{(top) Plot of the logarithm of the
calculated column densities of the cyanopolyynes in TMC-1 from
HC$_{5}$N to HC$_{11}$N for the indicated excitation
temperatures.(bottom) Plot of the decrements between successive
odd-numbered cyanopolyynes in TMC-1 for the indicated  excitation
temperatures.  
\label{fig3}}


\newpage


\begin{references}
\reference{krot1976} Alexander, A. J., Kroto, H. W. \& Walton,
D.R.M., 1976,  J. Mol. Spec., 62, 175
\reference{aver1980} Avery, L. W. 1980, in \em Interstellar
Molecules, \em IAU  Symposium No. 87, ed. B. H. Andrew, publ. (D.
Reidel, Dordrecht:Holland), p47
\reference{aver1992} Avery, L. W. et al. 1992, \apjs, 83, 363
\reference{bell1996} Bell, M. B. 1996, \apjl, 453, 773
\reference{bell1997} Bell, M. B. 1997, \pasp (in press, May 1997)
\reference{bell1992} Bell, M. B., Avery, L. W., MacLeod, J. M., \&
Matthews,  H. E. 1992, \apj, 400, 551
\reference{bell1993} Bell, M. B., Avery, L. W., \& Watson, J. K.
G. 1993, 
\apjs, 86, 211
\reference{bell1992a} Bell, M. B., Feldman, P. A., \& Avery, L. W.
1992, 
\apj, 396, 643
\reference{bell1997a} Bell, M. B., Feldman, P. A. \& Watson,
J.K.G. 1997,  (in preparation)
\reference{bell1982} Bell, M. B., Feldman, P. A., Kwok, S. \&
Matthews, H. E.  1982, \nat, 295, 389
\reference{bell1985} Bell, M. B. \& Matthews, H. E. 1985, \apjl,
291, L63
\reference{brot1978} Broten, N. W., Oka, T., Avery, L. W. \&
MacLeod, J. M.  1978, \apjl, 223, L105
\reference{chur1978} Churchwell, E., Winnewisser, G., and
Walmsley, C. M. 1978, \aap, 67, 139
\reference{dick1975} Dickman, R.L. 1975, \apj, 202, 50
\reference{krot1978} Kroto, H.W.  et al. 1978, \apjl, 219, L133
\reference{olan1988} Olano, C.A., Walmsley, C.M. and Wilson, T.L.
1988, \aap, 196, 194
\reference{snyd1986} Snyder, L. E., Dykstra, C.E., \& Bernholdt,
D. 1986, in 
\em Masers, Molecules and Star Forming Regions \em, ed. A. D.
Haschick  (Westford, MA: Haystack Obs. Press), p9
\reference{toll1981} T\"{o}lle, F., Ungerechts, H., Walmsley, C.
M.,  Winnewisser, G., \& Churchwell, E. 1981, \aap, 95, 143
\reference{trav1996} Travers, M. J., McCarthy, M. C., Kalmus, P.,
Gottlieb, C.  A., \& Thaddeus, P. 1996, \apjl, 469, L65. 
\end{references}
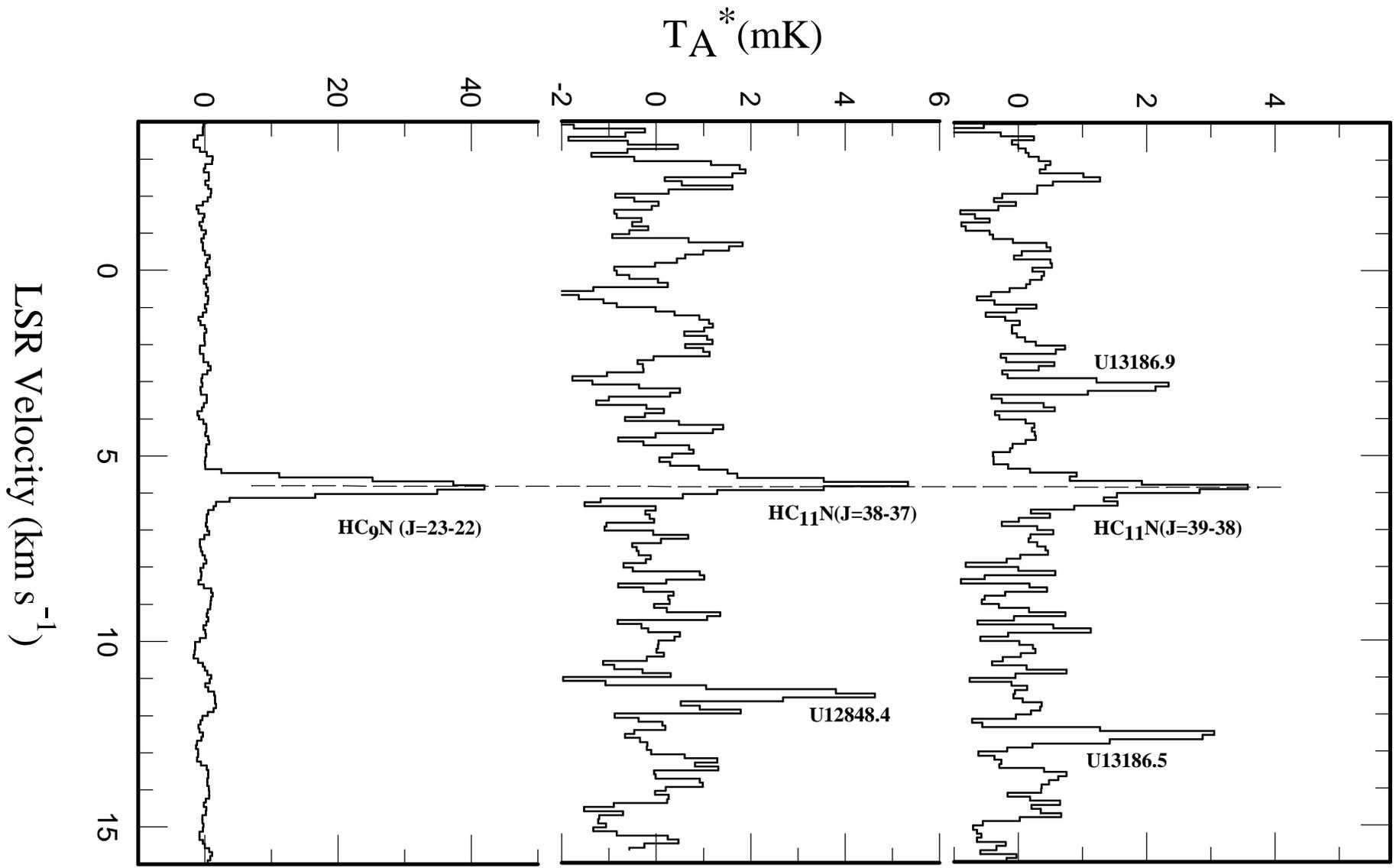
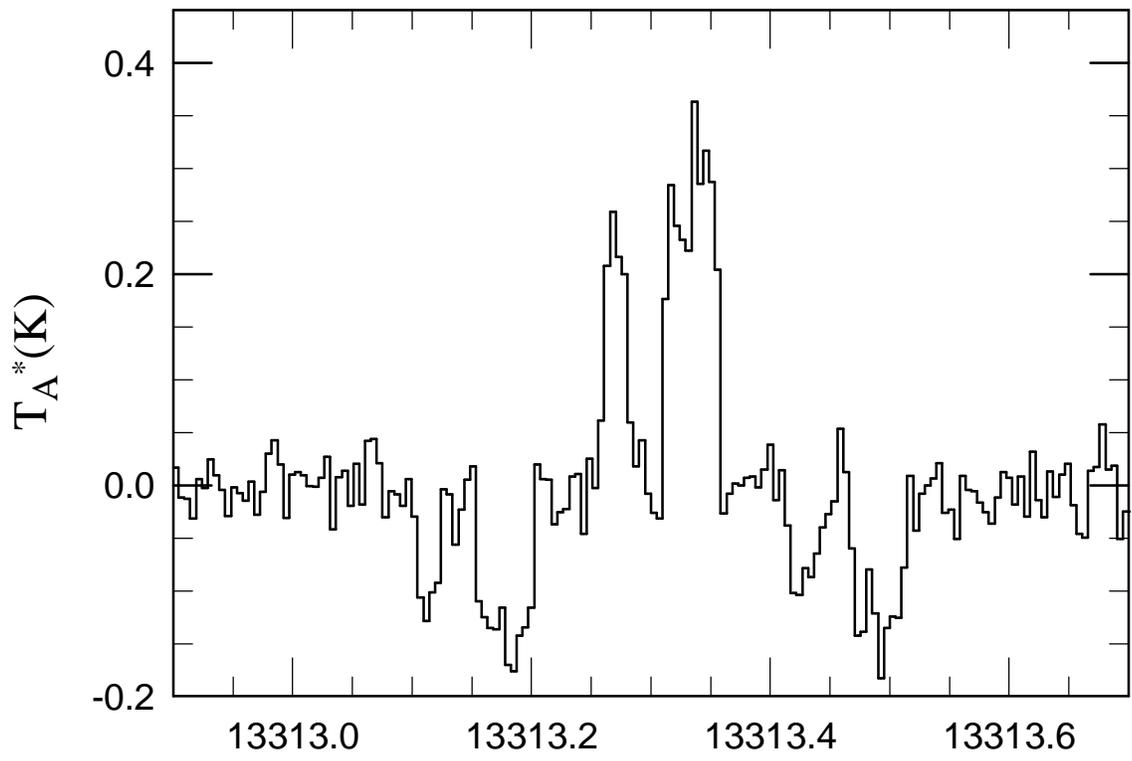
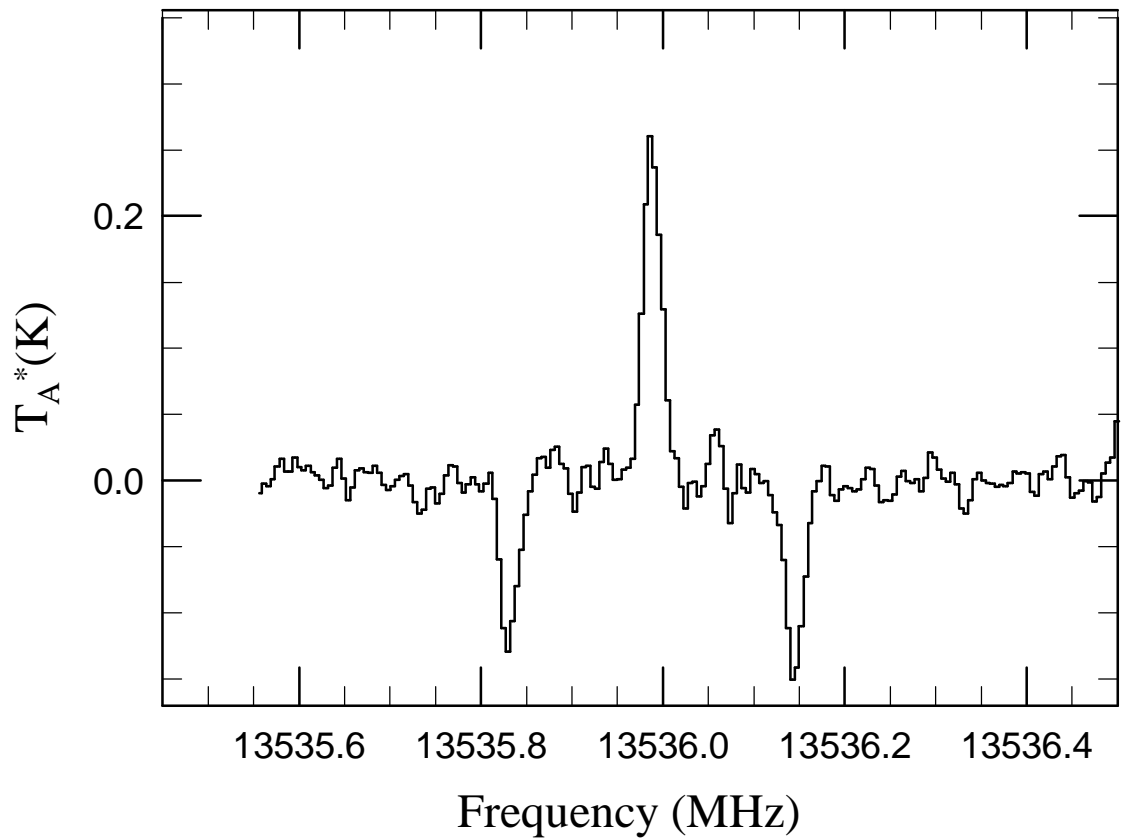
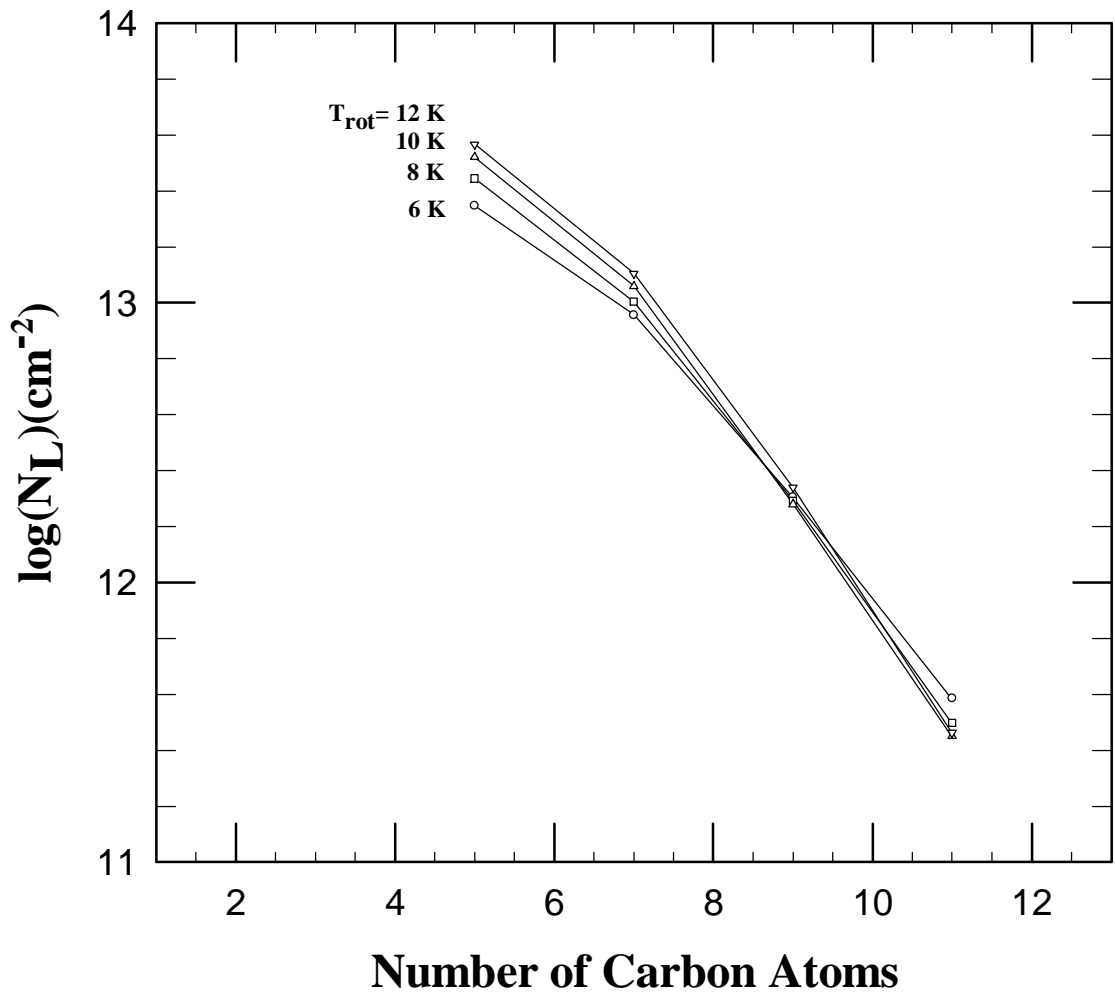
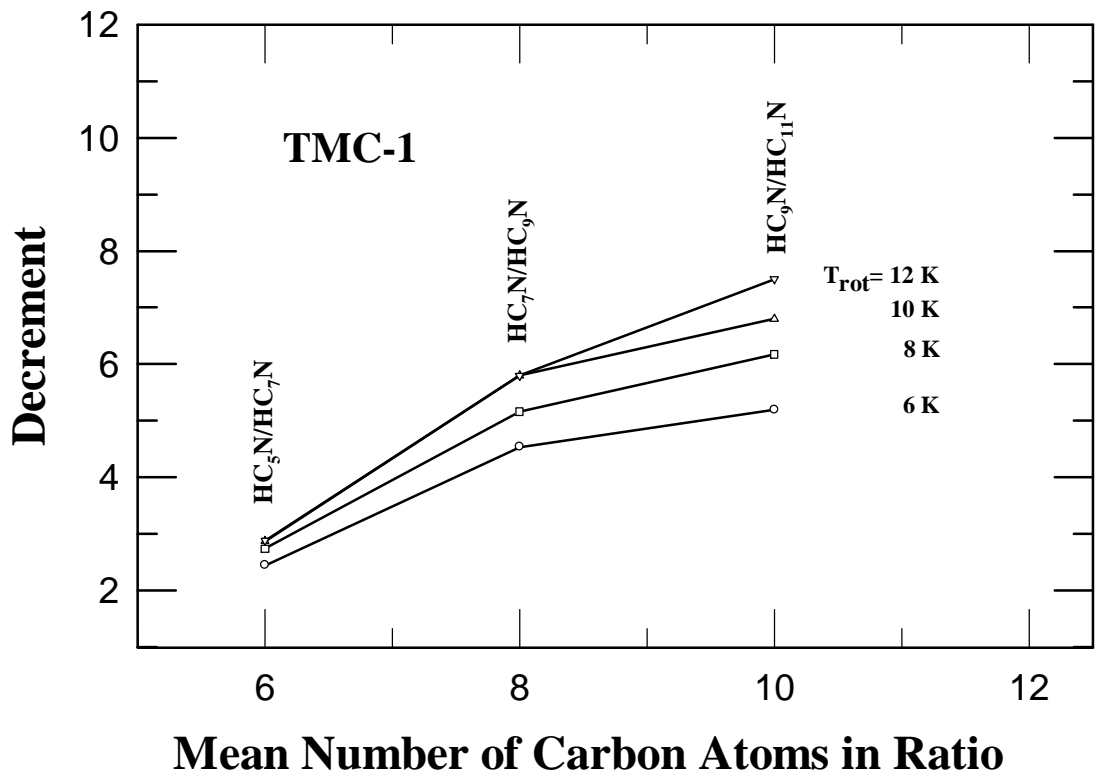
\end{document}